\documentclass{ws-mplb}
\usepackage{multirow}

\begin{document}

\graphicspath{{eps/}}

\newcommand{\atan}{{\rm Arctan}}
\newcommand{\af}{antiferromagnetic}
\newcommand{\const}{{\rm const}}
\newcommand{\half}{\frac{1}{2}}
\newcommand{\halfpi}{\frac{\pi}{2}}
\newcommand{\hb}{\bar{h}}
\newcommand{\id}{{\bf 1}}
\newcommand{\Lb}{\bar{L}}
\newcommand{\On}{{\rm O}(n)}
\newcommand{\qb}{\bar{q}}
\newcommand{\resp}{resp.}
\newcommand{\Rc}{\check{R}}
\newcommand{\rmd}{{\rm d}}
\newcommand{\TL}{{\rm TL}}
\newcommand{\ul}{\underline}
\newcommand{\Uq}{{\rm U}_q({\rm Sl}_2)}
\newcommand{\wh}{\widehat}
\newcommand{\wt}{\widetilde}
\newcommand{\Zbb}{\mathbb{Z}}

\newcommand{\myscript}[2]{\begin{array}{c}
    {\scriptstyle #1} \\
    {\scriptstyle #2}
  \end{array}}

\markboth{Yacine Ikhlef}{The antiferromagnetic Potts model}

%
\catchline{}{}{}{}{}
%

\title{The antiferromagnetic Potts model}

\author{\footnotesize Yacine Ikhlef}

\address{
  Section Math\'ematiques, Universit\'e de Gen\`eve, \\
  2-4 rue du Li\`evre, 1211 Gen\`eve, Switzerland \\
  yacine.ikhlef@unige.ch
}

\maketitle

\begin{history}
  \received{(Day Month Year)}
  \revised{(Day Month Year)}
\end{history}


\begin{abstract}
  We review the exact results on the various critical regimes of the
  antiferromagnetic $Q$-state Potts model. We focus on the Bethe Ansatz
  approach for generic $Q$, and describe in each case the effective
  degrees of freedom appearing in the continuum limit.
\end{abstract}


\section{Introduction}

The $Q$-state Potts model\cite{Potts}, defined in 1952, is a lattice model for classical
magnetism. Since the 1970's, it has been
used as a laboratory to develop theoretical methods of Statistical Mechanics,
such as Kramers-Wannier duality, Yang-Baxter
integrability, Bethe Ansatz\cite{Baxter-book},
Coulomb-gas formalism\cite{CG-Nienhuis,CG-Hubert-etal},
Conformal Field Theory\cite{CFT-book} (CFT). Most studies on the Potts model deal with
the {\it ferromagnetic} critical point, described by a Coulomb-Gas CFT or
minimal CFT at the rational values.
In contrast, the present review is concerned with the {\it \af}
(AF) regime of the Potts model, which also contains exactly solvable critical points.
Although it is based on the same simple lattice model,
this regime exhibits interesting physical features. Indeed, non-unitarity of the
equivalent vertex model allows a CFT with central charge $c>1$, and non-compact
degrees of freedom.

Let us introduce the subject with a short historical notice.
In 1982, Baxter first discovered\cite{Baxter-PAF} the location
of the AF/paramagnetic transition by a mapping to a
staggered integrable six-vertex (6V) model, and calculated the free
energy density. Later on, Saleur described\cite{Saleur90} the complete phase
diagram for $0 \leq Q \leq 4$, emphasing the role of the Beraha numbers $Q=4\cos^2 \frac{\pi}{t}$
for integer $t$. At these values, he gave\cite{Saleur91} the CFT description
of the AF/paramagnetic transition, relating it to the $\Zbb_k$ parafermion CFT\cite{Zk}. 
This transition corresponds to a staggered integrable model, but it
appeared\cite{AuYang-Perk92}
as a solution of the basic Yang-Baxter Equations (YBE) for the $(N_\alpha,N_\beta)$
Potts model.
A complete CFT description of this
transition for both generic and Beraha $Q$ was achieved\cite{JS-PAF,IJS-PAF},
especially through the study of the Bethe Ansatz equations: it consists in
one compact boson $\phi_1$ with $Q$-dependent radius and one non-compact boson
$\phi_2$. Besides the isotropic critical regimes mentioned so far, a mixed
ferro-{\af} anisotropic regime was defined and solved\cite{IJS-PAF-or}.
Its continuum limit consists in two compact bosons: $\phi_1$ with a $Q$-dependent
radius, and $\phi_2$ with a fixed radius, such that $\phi_2$ decomposes into
two Ising models.

This review is organised as follows. Section~\ref{sec:Potts} is a reminder on the
Potts model and its phase diagram for generic $0 \leq Q \leq 4$, and describes some
related problems relevant to Statistical Mechanics, but also Condensed Matter Theory.
There are three critical regimes for the Potts model with AF interactions.
In Section~\ref{sec:approach}, we summarize the methods used to study the
critical points of the Potts model for generic values of $Q$.
Sections~\ref{sec:BK}, \ref{sec:PAF}, \ref{sec:PAF-or} then apply these methods
respectively to the BK phase, the AF transition line and the anisotropic
critical regime. Section~\ref{sec:conclusion} concludes by mentioning some aspects
that are not treated in detail by this review, together with some interesting open
problems.

\section{The Potts model and related Physics problems}
\label{sec:Potts}

\subsection{Definitions and integrability properties}
\label{sec:def}

Let $\cal L$ be a planar lattice. The $Q$-state Potts model on $\cal L$ consists of spins
$S_j$ living on the vertices of $\cal L$, and taking the values $\{ 1,2, \dots, Q \}$. The Boltzmann weight
for a spin configuration is
\begin{equation}
  \label{eq:Potts-weight}
  W[ \{ S_j \} ] = \exp \left(
    J \sum_{\langle i,j \rangle} \delta_{S_i,S_j}
  \right) \,,
\end{equation}
where $\delta$ stands for the Kronecker symbol and the product is on all edges of $\cal L$.
A positive ({\resp} negative) value of $J$ favors configurations where neighboring sites
have equal ({\resp} distinct) spins. This defines the ferromagnetic and AF regimes.

\begin{figure}[th]
  \centerline{\psfig{file=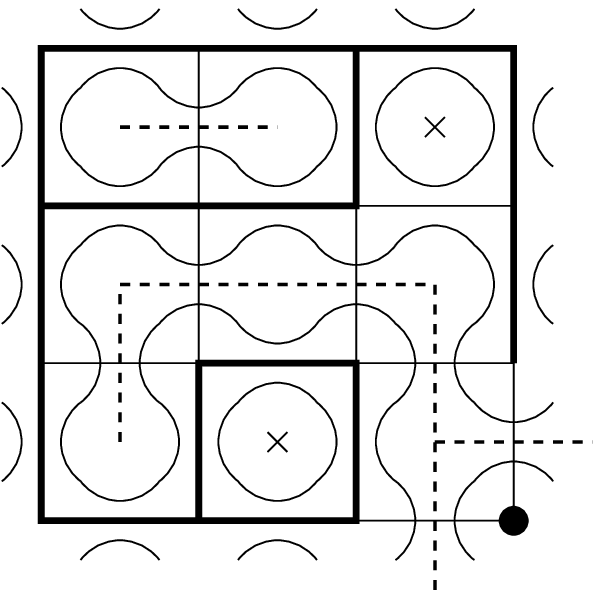, scale=0.5}}
  \vspace*{8pt}
  \caption{Example FK cluster configuration on the square lattice $\cal L$, and corresponding
    loop configuration. Direct ({\resp} dual) clusters are shown as thick full ({\resp} dotted)
    lines, and clusters consisting of a single site as dots ({\resp} crosses).}
  \label{fig:fk}
\end{figure}

The graphical expansion of the partition sum gives the Fortuin-Kasteleyn (FK) cluster
model\cite{FK} (see Fig~\ref{fig:fk}):
\begin{equation}
  \label{eq:FK}
  Z_{\rm Potts} = 
  \sum_{G \subseteq {\cal L}} Q^{C(G)} v^{\ell(G)}
  := Z_{\rm FK}({\cal L}, Q, v)  \,,
  \qquad v := e^J-1 \,,
\end{equation}
where $C(G)$ is the number of connected components (clusters) of the subgraph $G$, and
$\ell(G)$ is the number of edges in $G$. In this representation, $Q$ plays the role of
a real parameter, and we are mainly concerned with the regime $0 \leq Q \leq 4$.
Using Euler's relation, the partition sum can be expressed as
\begin{equation}
  \label{eq:Z-dual}
  Z_{\rm FK}({\cal L}, Q, v) =
  \const \times \sum_{G' \subseteq {\cal L'}} Q^{C(G')} \left( \frac{Q}{v} \right)^{\ell(G')}
  = \const \times Z_{\rm FK}({\cal L}', Q, Q/v) \,,
\end{equation}
where ${\cal L}'$ is the lattice dual to $\cal L$.
We refer to the transformation~\eqref{eq:Z-dual} as Kramers-Wannier duality.

The FK model can be, in turn, mapped to a loop model based on the Temperley-Lieb (TL) algebra\cite{TL}
(see Fig~\ref{fig:fk}).
The TL loop model lives on the medial lattice $\cal M$ consisting of the midpoints of
the original lattice $\cal L$. A cluster configuration defines uniquely a loop configuration,
by asking loops to separate direct and dual clusters. We have the relation
\begin{equation}
  \label{eq:Z-TL}
  \const \times Z_{\rm FK}({\cal L}, Q, v) = \sum_\text{loop config. $H$}
  n^{c(H)} x ^{X(H)} := Z_{\rm loop}  \,,
  \quad \begin{cases}
    n = \sqrt{Q} \\
    x = v / \sqrt{Q}
  \end{cases} \,,
\end{equation}
where $c(H)$ is the number of closed loops in $H$, and $X(H)$ is the number of medial sites
where the loops do not cross the original edge of $\cal L$.
Although the Potts model can be defined on any planar lattice, we restrict the
subsequent discussion to the square lattice for simplicity.

Consider the Potts model on a strip on width $L$ sites (with $L$ even), with
coupling constant $J_1$ for horizontal
edges, $J_2$ for vertical edges, and corresponding weights $x_1,x_2$ in the loop model.
The transfer matrix for loops can be built with the TL algebra $\TL_L(n)$. This algebra is generated
by the operators $e_j$ for $j=1, \dots, L-1$, satisfying:
\begin{equation}
  \label{eq:TL}
  \begin{array}{rcl}
    e_j^2 &=& n \ e_j \\
    e_j e_{j \pm 1} e_j &=& e_j \\
    e_j e_k &=& e_k e_j \quad \text{for $|j-k|>1$.}
  \end{array}
\end{equation}
The transfer matrix parallel to one axis of the original lattice $\cal L$
then reads
\begin{equation}
  \label{eq:T-parallel}
  T_\| = \left[
    \prod_\text{$j$ even} (\id + x_1 \ e_j)
  \right]
  \left[
    \prod_\text{$j$ odd} (x_2 \ \id + \ e_j)
  \right] \,.
\end{equation}
Our regime of interest is $0 \leq n \leq 2$.
We introduce the {\it crossing parameter} $\gamma$ and the parameter $t$, defined by
\begin{equation}
  \label{eq:def-gamma}
  n= \sqrt{Q} = 2\cos \gamma \,, \qquad 0 \leq \gamma \leq \frac{\pi}{2} \,,
  \qquad t:= \frac{\pi}{\gamma} \,.
\end{equation}
The algebra $\TL_L(n)$ provides a solution to the YBE, with one spectral parameter $u$:
\begin{equation}
  \label{eq:Rc}
  \Rc_{j,j+1}(u) = \sin(\gamma-u) \ \id + \sin(u) \ e_j \,.
\end{equation}

\subsection{Phase diagram}
\label{sec:phase-diagram}

\begin{figure}[th]
  \centerline{\psfig{file=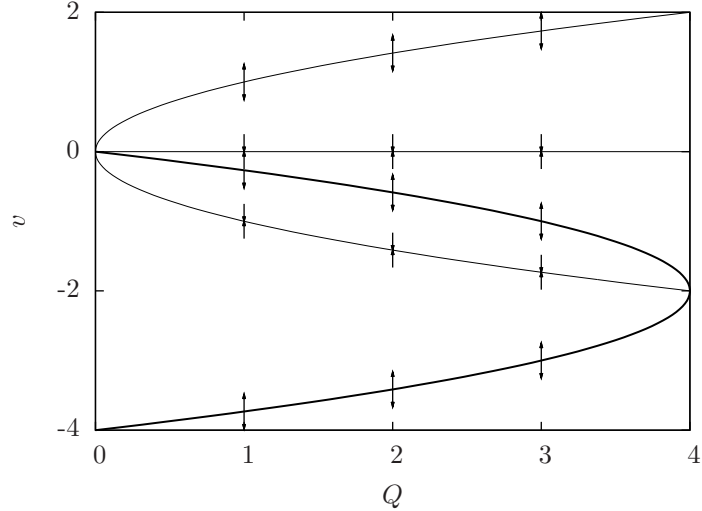, scale=1}}
  \vspace*{8pt}
  \caption{Phase diagram of the isotropic Potts model on the square lattice.
    The self-dual line $v=\pm \sqrt{Q}$ is shown as a thin line, and the AF critical line
    $v=-2\pm \sqrt{4-Q}$ as a thick line. The BK critical phase, governed by the self-dual
    branch $v=-\sqrt{Q}$, is enclosed in the thick line. Arrows represent the
    renormalization-group flows in the vicinity of fixed points.}
  \label{fig:phase-diag}
\end{figure}

Let us recall the location\cite{Baxter-book,Baxter-PAF} of the critical
points of the square-lattice Potts model.
\begin{itemize}
\item From \eqref{eq:T-parallel} and \eqref{eq:Rc}, we see that the self-dual line $x_1 x_2 = 1$
  corresponds to a homogeneous loop model with
  \begin{equation} \label{eq:x-sd}
    x_1 = \frac{\sin u}{\sin(\gamma-u)} \,, \qquad x_2 = \frac{1}{x_1} 
    \,.
  \end{equation}
\item The only other integrable case respecting the parity of sites in~\eqref{eq:T-parallel} is 
  when spectral parameters alternate between $u$ and $u+\frac{\pi}{2}$\cite{Baxter-PAF,IJS-PAF,IJS-PAF-or}:
  \begin{equation} \label{eq:x-stag}
    x_1 = \frac{\sin u}{\sin(\gamma-u)} \,, \qquad x_2 = -\frac{\cos(\gamma-u)}{\cos u}
    \,.
  \end{equation}
\end{itemize}
Each of these two cases contains two critical regimes, corresponding to
different values of the spectral parameters: see Table~\ref{table:regimes}.
The three isotropic critical regimes are shown in Fig.~\ref{fig:phase-diag}.

\begin{table}[pt]
  \tbl{The critical regimes of the Potts model on the square lattice.
  The last column gives the generic value of the central charge for the loop model.
  \label{table:regimes}}
  {\begin{tabular}{@{}ccccc@{}} \toprule
      \multirow{3}{*}{self-dual} &
      $0<u<\gamma$ & $x_1,x_2>0$ & ferromagnetic critical & $c=1-\frac{6}{t(t-1)}$ \\ \\ 
      & $\gamma<u<\pi/2$ & $x_1,x_2<0$ & BK phase & $c=1-\frac{6(t-1)^2}{t}$ \\ \colrule
      \multirow{3}{*}{staggered} &
      $\gamma<u<\pi/2$ & $x_1,x_2<0$ & AF critical & $c=2-\frac{6}{t}$ \\ \\ 
      & $0<u<\gamma$ & $x_1>0,x_2<0$ & anisotropic critical & $c=2-\frac{12}{t(t-2)}$ \\ \botrule
    \end{tabular} }
\end{table}

\subsection{Related problems of Statistical Mechanics and Condensed Matter Theory}
\label{sec:related}

Using the FK formulation, one can relate the Potts model to various statistical models through
exact equivalences.
\begin{itemize}

\item {\bf Percolation.} At $Q=1$, the FK model reduces to bond percolation on the lattice $\cal L$.

\item {\bf Uniform spanning trees.} In the limit $(Q \to 0, v=\sqrt{Q})$, using Euler's relation,
  \begin{equation}
    Q^{-\frac{1}{2}({\cal N}+1)} Z_{\rm FK}({\cal L}, Q, \sqrt{Q})
    \ \mathop{\longrightarrow}_{Q \to 0} \ \# \ \text{spanning trees on $\cal L$} \,,
  \end{equation}
  where $\cal N$ is the total number of sites in $\cal L$.
  So the continuum limit of the uniform spanning tree problem\cite{UST} is determined
  by the ferromagnetic Potts transition at $Q \to 0$.

\item {\bf Spanning forests.} In the vicinity of $Q=0$, one can define another
  limit: $(Q \to 0, v= Q/w)$, where $w$ is constant.
  Using the same arguments as above, we get the limit:
  \begin{equation}
    (Q/w)^{-({\cal N}+1)} Z_{\rm FK}({\cal L}, Q, Q/w)
    \ \mathop{\longrightarrow}_{Q \to 0} \ \sum_{G \in {\cal F}({\cal L})} w^{C(G)} \,,
  \end{equation}
  where ${\cal F}({\cal L})$ is the set of spanning forests\cite{forest} on $\cal L$, {\it i.e.}
  spanning subgraphs with
  no internal face. The parameter $w$ thus plays the role of a tree fugacity in the spanning
  forest problem. The phase diagram of the Potts model predicts\cite{forest-sigma} the behavior of
  spanning forest as a function of $w$: if $w>0$ or $w<-4$, the model is not critical
  (paramagnetic phase); if $-4<w<0$, the model is described by a Coulomb-Gas theory (BK phase);
  exactly at $w=-4$, the model is described by an ${\rm OSP}(2|2)$
  supersphere $\sigma$-model\cite{forest-sigma}.

\item {\bf Restricted Solid-On-Solid (RSOS) models.}
  At the Beraha numbers $\sqrt{Q} = 2 \cos \frac{\pi}{t}$ with
  integer $t \geq 4$, the TL
  generators can be represented as local operators in
  the $A_{t-1}$ RSOS height model\cite{RSOS1,RSOS2}.
  The latter consists of local variables $\{ h_j , j=1, \dots, L \}$ subject to the local constraints
  \begin{equation}
    h_j \in \{1, \dots, t-1 \} \,, \qquad |h_j-h_{j+1}|=1 \,.
  \end{equation}
  The operators $e_j$ are defined as
  \begin{equation}
    e_j |h_1 \dots h_L \rangle = \delta_{h_{j-1},h_{j+1}}
    \sum_{h' = h_{j+1} \pm 1}
    \frac{\sqrt{\sin \frac{\pi h_j}{t} \sin \frac{\pi h'}{t}}}{\sin \frac{\pi h_{j+1}}{t}}
    |h_1 \dots h' \dots h_L \rangle \,.
  \end{equation}
  The RSOS model may describe the statistics of a two-dimensional interface
  in three-dimensional space.
  
\end{itemize}

A second approach is to take the very anisotropic limit $u \to 0$, where the transfer
matrix\footnote{
  We consider here the transfer matrix parallel to one axis of the medial lattice $\cal M$.
} generates a local one-dimensional quantum Hamiltonian based on the $\TL_L(n)$ algebra\footnote{
  We use a periodic variant of $\TL_L(n)$, with $e_{j \pm L}:=e_j$.
},
and consider the spin-$\half$ representation. The self-dual~\eqref{eq:x-sd} and
staggered~\eqref{eq:x-stag} transfer matrices
respectively give the Hamiltonians\cite{Baxter-book,IJS-PAF}:
\begin{eqnarray}
  H_{\rm sd} &=& \pm \sum_{j=1}^{L} e_j \,, \label{eq:H-sd} \\
  H_{\rm stag} &=& \pm \sum_{j=1}^{L} (-n \ e_j + e_j e_{j+1} + e_{j+1} e_j) \label{eq:H-stag} \,.
\end{eqnarray}
The global minus sign in $H_{\rm sd}$ ({\resp} $H_{\rm stag}$) corresponds to the ferromagnetic ({\resp} AF)
critical point; the plus sign to the BK phase ({\resp} the anisotropic critical point).

The TL generators can be
represented as rescaled projectors
in the $\Uq$ quantum algebra\cite{q-groups}, where $q=e^{i\gamma}$. Consider a collection
of $L$ spin-$\half$ variables $\{\sigma_j\}$. 
In terms of the Pauli matrices $\sigma_j^{x,y,z}$,
the self-dual Hamiltonian describes an XXZ quantum spin chain\cite{Baxter-book}:
\begin{equation} \label{eq:XXZ}
  H_{\rm sd} = \pm \frac{1}{2} \sum_{j=1}^L
  \left(
    \sigma_j^x \sigma_{j+1}^x + \sigma_j^y \sigma_{j+1}^y
    - \cos \gamma \ \sigma_j^z \sigma_{j+1}^z
  \right) + \const \,.
\end{equation}
If we denote $P_{j,j+1}^{(0)}$ the projector of $(\sigma_j+\sigma_{j+1})$ onto total
spin zero, then the operators $e_j:= n P_{j,j+1}^{(0)}$ satisfy the $\TL_L(n)$ relations~\eqref{eq:TL}.
Furthermore, the projector of $(\sigma_j+\sigma_{j+1}+\sigma_{j+2})$ onto total
spin $\frac{3}{2}$ is given by
\begin{equation*}
  P_{j,j+1,j+2}^{(3/2)} = 1 + \frac{(e_j e_{j+1} + e_{j+1} e_j) - n(e_j+e_{j+1})}{n^2-1} \,.
\end{equation*}
In terms of the $\Uq$ projectors, the staggered Hamiltonian reads\cite{IJS-TL}
\begin{equation} \label{eq:H-stag2}
  H_{\rm stag} = \pm \sum_{j=1}^L \left[
    n^2 P_{j,j+1}^{(0)} + (n^2-1) P_{j,j+1,j+2}^{(3/2)}
  \right] \,.
\end{equation}
If the plus sign is chosen and $n>1$, the second term in~\eqref{eq:H-stag2} favors a totally
dimerized state (a product of $L/2$ $\Uq$ singlets), while the first term gives additional energy
to dimerized pairs of neighbours: the spin chain is thus subject to frustration.
So $H_{\rm stag}$ is an integrable, $q$-deformed generalization of the well-studied ${\rm SU}(2)$
frustrated spin chain\cite{MG,frust-dmrg,qMG}.

\section{General approach to the solution}
\label{sec:approach}

\subsection{Equivalent vertex model, twisted BC}

We consider the Potts model on the lattice {\cal L}, embedded in a cylinder.
Its exact solution is obtained through the sequence of mappings\cite{Baxter-book}:
\begin{equation*}
  (\text{Potts on $\cal L$})
  \longrightarrow (\text{FK on $\cal L$})
  \longrightarrow (\text{TL on $\cal M$})
  \longrightarrow (\text{6V on $\cal M$}).
\end{equation*}
The resulting 6V model is homogeneous for self-dual Potts, and staggered for
the AF and anisotropic critical lines.
The last mapping is done by orienting independently
each loop, and giving it a weight $e^{\pm i\gamma}$ according to its orientation, so that
the total weight for a loop is $n=2\cos \gamma$. The complex weight $e^{\pm i\gamma}$ is
distributed locally in the 6V model, by counting the local turns of each loop
fragment\cite{Baxter-book}. However, through this mapping, the loops which
wind around the cylinder ({\it non-contractible loops}) recieve an incorrect weight
$\wt{n}=2$, because their total rotation angle is zero in both orientations.
To correct this, we consider the 6V with a seam along the axis of the cylinder: an arrow
which crosses the seam from left to right ({\resp} from right to left) gets a factor $e^{i\varphi}$
({\resp} $e^{-i\varphi}$). Non-contractible loops then get a weight
\begin{equation}
  \wt{n}=2\cos \varphi \,.
\end{equation}
The Potts ground state corresponds to $\varphi=\gamma$, but it can be useful to consider other
values of $\varphi$, {\it e.g.}, to compute the magnetic exponent of the Potts model
(see below).

\subsection{Analysis of the BAE}

The integrable 6V is solvable by Bethe Ansatz.
In all the cases we consider here, the Bethe roots sit on horizontal line(s) in the
complex plane, and are described by a root density $\rho$ in the large-$L$ limit.
One may extract from the BAE two types of data which are
relevant to the associated critical theory:
\begin{itemize}
\item The physical properties of elementary excitations (particle-hole excitations
  in the root distribution) are derived easily by Fourier transform. This leads to
  the scattering theory of these quasi-particles both
  at criticality and in the vicinity of the critical point.
\item The finite-size corrections to the energies allow us to relate the lattice model to
  CFT, where the energies and momenta are expected to behave as
  \begin{eqnarray}
    E_0 &\simeq& e_\infty L - \frac{\pi v c}{6L} \,, \label{eq:E0-CFT} \\
    E_{h,\hb} &\simeq& E_0 + \frac{2\pi v}{L}(h + \hb) \,, \label{eq:E-CFT}  \\
    k_{h,\hb} &\equiv& \frac{2\pi}{L} (h-\hb) \ \mod 2\pi \,,
  \end{eqnarray}
  where $E_0$ is the ground-state energy, $e_\infty$ is the energy density, $v$ is the Fermi velocity,
  $c$ is the central charge and $E_{h,\hb},k_{h,\hb}$ are the energy and total momentum
  associated to the primary state with dimensions $(h,\hb)$. The computation of the finite-size
  corrections involves the Wiener-Hopf technique, described in this context by Yang
  and Yang\cite{YY}.
\end{itemize}

\subsection{Numerical transfer-matrix computations}

Analytical results may be checked
numerically by computing the dominant eigenvalues of the transfer matrix in
various sectors or, equivalently, the lowest eigenvalues of the Hamiltonian.
Through \eqref{eq:E0-CFT}-\eqref{eq:E-CFT}, the eigenvalues provide numerical
estimates for the central charge and conformal dimensions.

Using the TL formulation, one usually computes systems of up to $L=20$ sites,
and the precision obtained for the exponents is about $10^{-3}$ (except in the presence
of logarithmic corrections, typically at $\gamma \to 0$).

\section{The Berker-Kadanoff phase}
\label{sec:BK}

\subsection{Relation to the XXZ spin chain}

The two branches of the self-dual line~\eqref{eq:x-sd} are related to each other
by $(x_1,x_2) \to (-x_1,-x_2)$, or equivalently by $\sqrt{Q} \to -\sqrt{Q}$ (see~\eqref{eq:TL}).
We denote $\mu:=\pi-\gamma$, so that $-\sqrt{Q}=2\cos \mu$.
In the very anisotropic limit $u \to 0$, the BK critical line is thus equivalent to
an XXZ spin chain with twisted BC:
\begin{equation}
  H_{\rm XXZ} := -\half \sum_{j=1}^L \left(
    \sigma_j^x \sigma_{j+1}^x + \sigma_j^y \sigma_{j+1}^y
    + \Delta \ \sigma_j^z \sigma_{j+1}^z
  \right) \,,
  \quad
  \sigma_{L+1}^{\pm} := e^{\pm 2 i \varphi} \sigma_1^{\pm} \,,
\end{equation}
where
\begin{equation}
  \Delta = -\cos \mu \,,
  \qquad
  \varphi = \mu \,.
\end{equation}
In the present Section, we will then recall Bethe-Ansatz results on XXZ, and
their application to the study of the BK phase for generic $Q$.

On each site of the XXZ chain, we consider an up spin as an empty site,
and a down spin as a particle. The Bethe Ansatz Equations and
eigenvalues for $r$ particles read:
\begin{eqnarray}
  \left[
    \frac{\sinh(\frac{i\mu}{2}-\lambda_j)}
    {\sinh(\frac{i\mu}{2}+\lambda_j)}
  \right]^L &=& (-1)^{r-1} e^{2i\varphi} \prod_{\ell=1}^r
  \frac{\sinh(i\mu-\lambda_j+\lambda_\ell)}
  {\sinh(i\mu+\lambda_j-\lambda_\ell)} \,,
  \label{eq:BAE} \\
  E &=& \sum_{j=1}^r \epsilon(\lambda_j)
  = \sum_{j=1}^r \frac{-2\sin^2 \mu}{\cosh \lambda_j - \cos \mu} \,.
  \label{eq:E-BA}
\end{eqnarray}

\subsection{The ground state and the quasi-particle picture}

We take the logarithmic form of the BAE:
\begin{equation}
  L k(\lambda_j) = 2\pi I_j + 2\varphi - \sum_{\ell=1}^r \Theta(\lambda_j-\lambda_\ell) \,,
\end{equation}
where
\begin{equation}
  k(\lambda) := -i \log \frac{\sinh(i\frac{\mu}{2}-\lambda)}
  {\sinh(i\frac{\mu}{2}+\lambda)} \,,
  \qquad
  \Theta(\lambda) := -i \log \frac{\sinh(i\mu+\lambda)}
  {\sinh(i\mu-\lambda)} \,,
\end{equation}
and the ``Bethe integers'' $I_j$ are half-odd integers ({\resp} integers) if $r$ is
even ({\resp} odd). In the large-$L$ limit, the Bethe roots $\lambda_j$ are described by
a root density $\rho(\lambda)$, satisfying the linear integral equation
\begin{equation} \label{eq:lieb}
  k'(\lambda) = 2\pi \rho(\lambda)
  - \int_{\Lambda_-}^{\Lambda_+} \rmd\nu \ \rho(\nu) K(\lambda-\nu) \,,
\end{equation}
where the kernel $K$ is given by $K:=\Theta'$, and $\Lambda_\pm$ are the extremal values of
the $\lambda_j$'s.

The ground state for $\varphi=0$ corresponds to a symmetric root distribution
with $r=L/2$ and $\Lambda_\pm = \pm \infty$, so \eqref{eq:lieb} becomes solvable by Fourier transform.
We use the notations
\begin{equation}
  \wh{f}(\omega) := \int_{-\infty}^{+\infty} \rmd\lambda \ f(\lambda) e^{i\omega \lambda} \,,
  \qquad
  (f \star g)(\lambda) := \int_{-\infty}^{+\infty} \rmd\nu \ f(\nu) g(\lambda-\nu) \,.
\end{equation}
The ground-state solution is 
\begin{equation}
  \rho_0(\lambda) = \frac{1}{2\mu \cosh \frac{\pi \lambda}{\mu}} \,,
  \qquad
  \wh{\rho}_0(\omega) =  \frac{1}{2 \cosh \frac{\mu \omega}{2}} \,.
\end{equation}
Any physical quantity is given by an integral of the form
\begin{equation}
  A(\rho) = \int_{\Lambda_-}^{\Lambda_+} \rmd\lambda \ \rho(\lambda) a(\lambda) \,,
\end{equation}
where, {\it e.g.}, $a=1$ for the particle density, $a=k$ for the total momentum, {\it etc}.
A hole at position $\lambda_h$ in the root distribution produces a variation
in $A$, given by the dressed quantity\cite{QISM} $\wt{a}$:
\begin{equation}
  A(\rho)-A(\rho_0) = - \frac{1}{L}
  \left[ \left( 1-\frac{K}{2\pi} \right)^{-1} \star a \right](\lambda_h)
  := -\frac{1}{L} \wt{a}(\lambda_h) \,.
\end{equation}
In particular, the dispersion relation of the quasi-particles is given
by:
\begin{equation} \label{eq:dispersion}
  \wt{k}(\lambda) = 2 \atan \left( \tanh \frac{\pi \lambda}{2\mu} \right) \,,
  \qquad
  \wt{\epsilon}(\lambda) = -\frac{\pi \sin \mu}{\mu \cosh \frac{\pi \lambda}{\mu}} \,.
\end{equation}
Close to the Fermi levels $\wt{k}_F = \pm \frac{\pi}{2}$, this becomes a linear relation:
\begin{equation}
  \wt{\epsilon} \simeq v |\wt{k}-\wt{k}_F| \,,
  \qquad
  v = \frac{\pi \sin \mu}{\mu} \,.
\end{equation}
To complete the $S$-matrix picture, one defines the dressed scattering amplitudes
as follows. Consider a root distribution with particles of density $\rho$ and holes
of density $\rho_h$. The BAE then read
\begin{equation}
  k' = 2\pi(\rho+\rho_h) - K \star \rho \,.
\end{equation}
Introducing $\Phi := (2\pi-K)^{-1} \star (-2\pi K)$, we express the BAE as
\begin{equation}
  \wt{k}' = 2\pi(\rho+\rho_h) - \Phi \star \rho_h \,.
\end{equation}
In the above form, the BAE describe the scattering of holes with momentum
$\wt{k}$ and scattering amplitude
\begin{eqnarray}
  S(\lambda,\lambda') &=& \exp \left[
    i \int_0^{\lambda-\lambda'} \rmd\nu \ \Phi(\nu)
  \right] \\
  &=& \exp \left[
    -i \int_{-\infty}^{+\infty} \frac{\rmd\omega}{\omega}
    \ \frac{\sinh \left(\mu-\frac{\pi}{2}\right)\omega\ \sin(\lambda-\lambda') \omega} 
    {2 \cosh \frac{\mu\omega}{\pi} \ \sinh \frac{\pi-\mu}{2} \omega}
  \right] \,.
\end{eqnarray}

\subsection{CFT spectrum}
\label{sec:CFT-XXZ}

In this paragraph, we use the BAE to derive the CFT associated to the {\it untwisted}
($\varphi=0$) spin chain. In the next paragraph, we will then reintroduce the twist to
compute correctly the critical exponents in the Potts model.

As explained above, the CFT spectrum is obtained through the finite-size corrections
to the energies. The central charge is given by the ground-state energy $E_0$.
In this case, the corrections are easily computed\cite{QISM} using the Euler-Maclaurin
formula. This gives
\begin{equation}
  E_0 \simeq L \int_{-\infty}^{+\infty} \rmd\lambda \ \rho_0(\lambda) \epsilon(\lambda)
  - \frac{\pi v}{6L} \,,
\end{equation}
and so the central charge is $c=1$.

The states corresponding to primary operators in the CFT are combinations of
``magnetic'' and ``electric'' excitations. A magnetic excitation of charge $m \in \Zbb$
consists in removing $m$ roots from the ground state, and arranging the Bethe integers $I_j$
symmetrically around zero. An electric excitation of charge $e \in \Zbb$ is a global shift of
all Bethe integers: $I_j \to I_j-e$. The associated energy corrections may be computed by
the Wiener-Hopf technique\cite{YY}. This results in the expression
\begin{equation}
  E_{e,m} \simeq E_0 +\frac{2\pi v}{L} \left(
    \frac{g m^2}{2} + \frac{e^2}{2g}
  \right) \,,
  \qquad g=\frac{\pi-\mu}{\pi} \,.
\end{equation}
On the other hand, the total momentum for the $(e,m)$ excitation is
\begin{equation}
  k_{e,m} = \frac{2\pi}{L} \sum_{j=1}^r I_j \equiv \frac{2\pi e m}{L} \ \mod 2\pi \,.
\end{equation}
From $E_{e,m}$ and $k_{e,m}$, we deduce the conformal dimensions:
\begin{equation} \label{eq:h-em}
  h_{e,m} = \frac{1}{4} \left(
    \frac{e}{\sqrt{g}} + m\sqrt{g}
  \right)^2 \,,
  \qquad
  \hb_{e,m} = \frac{1}{4} \left(
    \frac{e}{\sqrt{g}} - m\sqrt{g}
  \right)^2 \,,
  \qquad
  g=\frac{\pi-\mu}{\pi} \,.
\end{equation}
We call ${\cal O}_{e,m}$ the corresponding primary operator in the CFT.
One can also identify the Bethe root configurations associated to the descendants
of ${\cal O}_{e,m}$: the action of the Virasoro generator $L_{-n}$ ({\resp} $\Lb_{-n}$) amounts
to shifting the highest ({\resp} lowest) Bethe integer by $n$ ({\resp} $-n$).

The central charge $c=1$, the primary dimensions~\eqref{eq:h-em} and the Virosoro
descendants obtained by this analysis of the BAE correspond to a
compactified boson (or Coulomb gas) CFT with action\cite{CG-Nienhuis,CG-Hubert-etal}
\begin{equation} \label{eq:action-CG}
  {\cal A} = \frac{g}{4\pi} \int \rmd^2 x \ (\nabla \phi)^2 \,,
  \qquad
  \phi \equiv \phi+2\pi \,.
\end{equation}
In this theory, a magnetic excitation is a dislocation defect of amplitude $2\pi m$, whereas
an electric excitation corresponds to the insertion of the vertex operator $\exp(ie\phi)$.

The continuum partition sum on a torus of modular ratio $\tau$ has the form:
\begin{equation}
  Z_{\rm 6V} = \frac{1}{|\eta(q)|^2} \sum_{e,m \in \Zbb} q^{h_{em}} \qb^{\hb_{em}} \,,
  \qquad q:=e^{2i\pi \tau} \,,
\end{equation}
where $\eta$ is the Dedekind function.

\subsection{Effective central charge and critical exponents}
\label{sec:exponents-BK}

In the presence of a twist $\varphi$, the conformal dimensions \eqref{eq:h-em}
are modified by the change of electric charges
\begin{equation}
  e \to e+\frac{\varphi}{\pi} \,.
\end{equation}
The effective central charge and critical exponents for the Potts model
in the BK phase are obtained by considering the appropriate sectors for $m$ and $\varphi$
in the 6V model.

The Potts ground state has charges $(e_0=\mu/\pi, m=0)$, and hence the effective central charge
for the Potts model is
\begin{equation}
  c_{\rm eff} = 1 - \frac{6(1-g)^2}{g} = 1-\frac{6(t-1)^2}{t} \,.
\end{equation}

The magnetic, $\ell$-leg watermelon and thermal exponents $X_H$, $X_\ell$ and $X_T$
are defined through the relations
\begin{eqnarray}
  G_H(r,r') &:=&
  \left\langle \delta_{S_r,S_{r'}} - \frac{1}{Q} \right\rangle
  \simeq |r-r'|^{-2X_H} \,, \\
  G_\ell(r,r') &:=&
  \mathbb{P} \left[ \text{$r,r'$ are connected by $\ell$ strands}
  \right]
  \simeq |r-r'|^{-2X_\ell} \,, \\
  \xi &\simeq& |v-v_c|^{-\nu} \,, \quad X_T:=2-\frac{1}{\nu} \,,
\end{eqnarray}
where $\xi$ is the correlation length and $v_c=-\sqrt{Q}$ is the RG fixed point for the BK phase.
\begin{itemize}

\item The magnetic correlation function may be written as
  \begin{equation}
    G_H(r,r')=(1-Q^{-1}) \frac{Z_H(r,r')}{Z} \,,
  \end{equation}
  where $Z_H(r,r')$ is the partition sum of FK configurations where $r$ and $r'$ belong to
  the same cluster. In the cylinder geometry, we put $r$ and $r'$ at the ends of the cylinder,
  and the constraint for $Z_H$ is equivalent to forbidding any non-contractible loop: this is
  easily implemented by setting $\varphi=\pi/2$, so that $\wt{n}=0$. Hence, the magnetic exponent is
  \begin{equation}
    X_H = \frac{(1/2)^2}{2g} - \frac{e_0^2}{2g} = \frac{1/4-(1-g)^2}{2g} \,.
  \end{equation}

\item The watermelon exponents correspond to higher magnetization sectors. Note that $\ell$
  has to be even when $L$ is even. Exponent $X_\ell$ corresponds to the charges $(e=0,m=\ell/2)$,
  and hence:
  \begin{equation}
    X_\ell = \frac{g \ell^2}{8} - \frac{(1-g)^2}{2g} \,.
  \end{equation}

\item The thermal exponent $X_T$ corresponds to an irrelevant perturbation around $v=-\sqrt{Q}$
  (see Fig.~\ref{fig:phase-diag}), and thus we expect $X_T \geq 2$. In the sector $m=0$,
  non-contractible loops are allowed, and the twist must be of the form $\varphi=\pi(e_0+p)$, with
  integer $p$. This leads to the series of exponents
  \begin{equation}
    X_T(p) = \frac{(e_0+p)^2 - e_0^2}{2g} \,.
  \end{equation}
  The two lowest exponents in this series correspond to the values $p=-2,1$:
  \begin{equation}
    X_T = 2 \,, \qquad
    X'_T = \frac{3}{2g} - 1 \geq 2
  \end{equation}

\end{itemize}

\section{The {\af} critical line}
\label{sec:PAF}

\subsection{The $\Zbb_2$ staggered vertex model}
\begin{figure}[th]
  \centerline{\psfig{file=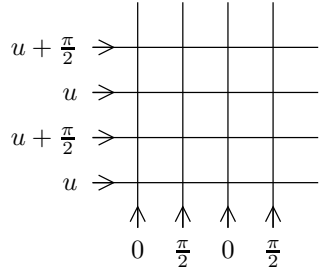, scale=1}}
  \vspace*{8pt}
  \caption{Spectral parameters of the $\Zbb_2$ staggered vertex model.}
  \label{fig:stag}
\end{figure}

The {\af} critical line is mapped to the staggered 6V model with an alternation of
spectral parameters, as shown in Fig.~\ref{fig:stag}. We consider this model on a cylinder of
circumference $L=2N$ sites. Since it is built on the TL algebra, the transfer matrix
has $\Uq$ invariance. Moreover, the YBE imply an additional symmetry, expressed by the
``$\Zbb_2$ conjugation operator'' $C$:
\begin{equation}
  C := \prod_{j=1}^N \left(-\frac{\Rc(\frac{\pi}{2})}{\cos \gamma} \right) \,.
\end{equation}
The operator $C$ satisfies $C^2=\id$ and commutes with the transfer matrix. As a consequence,
the eigenstates may be labelled according to this $\Zbb_2$ symmetry.

\subsection{Bethe Ansatz Equations}

The BAE and energies for the $\Zbb_2$ staggered model read, for even $N$:
\begin{equation}
  \left[
    \frac{\sinh(i\gamma+\alpha_j)}
    {\sinh(i\gamma-\alpha_j)}
  \right]^N
  = (-1)^{r-1} e^{2i\varphi} \prod_{\ell=1}^r
  \frac{\sinh \half(2i\gamma+\alpha_j-\alpha_\ell)}
  {\sinh \half(2i\gamma-\alpha_j+\alpha_\ell)} \,,
\end{equation}
\begin{equation} \label{eq:E-PAF}
  E = \sum_{j=1}^r \frac{2\sin^2 2\gamma}
  {\cosh 2\alpha_j - \cos 2\gamma} \,.
\end{equation}
The Bethe roots $\alpha_j$ are individually defined {\it modulo} $2i\pi$,
but the BAE are invariant under a global shift $\alpha_j \to \alpha_j+i\pi$.
This reflects the $\Zbb_2$ invariance discussed above, and in fact, one can
prove\cite{IJS-PAF} that the conjugation operator $C$ acts as follows on the Bethe
eigenstates $|\psi(\alpha_1,\dots,\alpha_r)\rangle$:
\begin{equation} \label{eq:C-psi}
  C |\psi(\alpha_1,\dots,\alpha_r)\rangle
  = \const \times |\psi(\alpha_1+i\pi,\dots,\alpha_r+i\pi)\rangle \,.
\end{equation}

The expression~\eqref{eq:E-PAF} for the energy suggests that Bethe roots
$\alpha_j$ lie on the lines ${\rm Im}\ \alpha = \pm \frac{i\pi}{2}$, so that their contribution
to the energy is negative. Hence, we divide the roots into
two sets
\begin{equation}
  \alpha_j^0= \lambda_j^0 + \frac{i\pi}{2} \,,
  \qquad
  \alpha_j^1= \lambda_j^1 - \frac{i\pi}{2} \,.
\end{equation}
We denote $r^a$ the number of roots $\lambda_j^a$.
In terms of the $\lambda_j^a$, the BAE and energies read
\begin{equation}
  \exp \left[i N k(\lambda_j^a) \right] = (-1)^{r^a-1} e^{2i\varphi}
  \prod_{b=0,1} \prod_{\ell=1}^{r^b}
  \exp \left[-i \Theta^{a-b}(\lambda_j^a-\lambda_\ell^b) \right] \,,
\end{equation}
\begin{equation}
  E = \sum_{a=0,1} \sum_{j=1}^{r^a} \epsilon(\lambda_j^a)
  = \sum_{a=0,1} \sum_{j=1}^{r^a} 
  \frac{-2\sin^2 2\gamma}{\cosh 2\lambda_j^a + \cos 2\gamma}
  \,,
\end{equation}
where:
\begin{equation}
  k(\lambda) := -i \log \frac{\cosh(i\gamma+\lambda)}
    {\cosh(i\gamma-\lambda)} \,,
\end{equation}
\begin{equation}
  \Theta^0(\lambda) := -i \log \frac{\sinh (i\gamma-\frac{\lambda}{2})}
  {\sinh (i\gamma+\frac{\lambda}{2})} \,,
  \qquad
  \Theta^{\pm 1}(\lambda) := -i \log \frac{\cosh (i\gamma-\frac{\lambda}{2})}
  {\cosh (i\gamma+\frac{\lambda}{2})} \,.
\end{equation}
In these notations, the $\Zbb_2$ conjugation~\eqref{eq:C-psi} exchanges the indices $a=0,1$.

\subsection{XXZ states}
\label{sec:XXZ-states}

If we denote $\mu=\pi-2\gamma$, then we have the identities
\begin{eqnarray}
  k(\lambda) &=& -i \log \frac{\sinh(\frac{i\mu}{2}-\lambda)}
  {\sinh(\frac{i\mu}{2}+\lambda)} \,,
  \label{eq:k-XXZ} \\
  (\Theta^0+\Theta^{\pm 1})(\lambda) &=&
  -i \log \frac{\sinh(i\mu+\lambda)}{\sinh(i\mu-\lambda)} \,,
  \label{eq:Theta-XXZ} \\
  \epsilon(\lambda) &=& \frac{-2\sin^2 \mu}{\cosh 2\lambda - \cos \mu}
  \label{eq:eps-XXZ} \,.
\end{eqnarray}
Hence, if the sets $\{ \lambda_j^0\}$ and $\{ \lambda_j^1\}$ are identical, the BAE and energies
are equivalent to those of the XXZ spin chain with parameter $\mu$ and twist $\varphi$.

\subsection{Ground state and quasi-particles}

The discussion of Section~\ref{sec:BK} on the large-$L$ limit of the BAE
can be adapted to the $\Zbb_2$ staggered model, except we now have two coupled
sets of BAE satisfying $\Zbb_2$ symmetry.
The logarithmic form of the BAE is
\begin{eqnarray}
  N k(\lambda_j^0) &=& 2\pi I^0_j + 2\varphi
  - \sum_{\ell=1}^{r^0} \Theta^0(\lambda_j^0-\lambda_\ell^0)
  - \sum_{\ell=1}^{r^1} \Theta^{-1}(\lambda_j^0-\lambda_\ell^1) \,, \\
  N k(\lambda_j^1) &=& 2\pi I^1_j + 2\varphi
  - \sum_{\ell=1}^{r^0} \Theta^1(\lambda_j^1-\lambda_\ell^0)
  - \sum_{\ell=1}^{r^1} \Theta^0(\lambda_j^1-\lambda_\ell^1) \,,
\end{eqnarray}
where the $I_j^a$ are half-odd integers ({\resp} integers) if $r^a$ is even ({\resp} odd).
In the large-$N$ limit, the roots lie on the intervals $[\Lambda_-^0,\Lambda_+^0]$ and
$[\Lambda_-^1,\Lambda_+^1]$, and the BAE become
\begin{equation}
  k'(\lambda) = 2\pi \rho^a(\lambda)
  - \sum_{b=0,1} \int_{\Lambda_-^b}^{\Lambda_+^b} \rmd \nu
  \ \rho^b(\nu) K^{a-b}(\lambda-\nu) \,,
\end{equation}
where $K^a := (\Theta^a)'$.
The ground state is a particular case of the XXZ states described in Section~\ref{sec:XXZ-states},
and hence the ground-state root densities are
\begin{equation}
  \rho^0(\lambda) = \rho^1(\lambda) = \rho_0(\lambda) =
  \frac{1}{2 \mu \cosh \frac{\pi \lambda}{\mu}} \,.
\end{equation}
A physical quantity reads
\begin{equation}
  A(\rho^0,\rho^1) = \int_{\Lambda_-^0}^{\Lambda_+^0}
  \rmd \lambda \ \rho^0(\lambda) a(\lambda)
  + \int_{\Lambda_-^1}^{\Lambda_+^1}
  \rmd \lambda \ \rho^1(\lambda) a(\lambda)
\end{equation}
We introduce the linear combinations $K^\pm := K^0 \pm K^1$. The presence of a hole $\lambda_h^0$
in the root distribution $\{ \lambda_j^0 \}$ results in the change:
\begin{equation}
  A(\rho^0, \rho^1) - A(\rho_0,\rho_0)
  = -\frac{1}{N} \left[
    \left( 1-\frac{K^+}{2\pi} \right)^{-1} \star a
  \right](\lambda_h^0) := \wt{a}(\lambda_h^0) \,.
\end{equation}
So, from \eqref{eq:k-XXZ}--\eqref{eq:eps-XXZ}, the elementary excitations have the same
dispersion relation~\eqref{eq:dispersion} as in the XXZ chain. The important
difference with XXZ is that there are now two kinds $(0,1)$ of quasi-particles,
and their dressed scattering amplitudes depend on the type of quasi-particles.
The BAE in the presence of holes read
\begin{equation}
  \left\{ \begin{array}{rcl}
      k' &=& 2\pi (\rho^0 + \rho_h^0) - K^0 \star \rho^0 - K^{-1} \star \rho^1 \\
      k' &=& 2\pi (\rho^1 + \rho_h^1) - K^1 \star \rho^0 - K^0 \star \rho^1 \,.
    \end{array} \right.
\end{equation}
We put them in ``diagonal form'':
\begin{equation}
  \left\{ \begin{array}{rcl}
      2k' &=& 2\pi (\rho^+ + \rho_h^+) - K^+ \star \rho^+ \\
      0 &=& 2\pi (\rho^- + \rho_h^-) - K^- \star \rho^- \,.
    \end{array} \right.
\end{equation}
where $\rho^\pm=\rho^0 \pm \rho^1$.
We can then express the BAE in terms of the $\rho_h^\pm$:
\begin{equation}
  \left\{ \begin{array}{rcl}
      2 \wt{k}' &=& 2\pi (\rho^+ + \rho_h^+) - \Phi^+ \star \rho_h^+ \\
      0 &=& 2\pi (\rho^- + \rho_h^-) - \Phi^- \star \rho_h^- \,.
    \end{array} \right.
\end{equation}
where $\Phi^\pm := (2\pi-K^\pm)^{-1} \star (-2\pi K^\pm)$. Finally, we get the dressed BAE:
\begin{equation}
  \wt{k}' = 2\pi (\rho^a+\rho_h^a) - \sum_{b=0,1} \Phi^{a-b} \star \rho_h^b \,,
\end{equation}
where $\Phi^0:= \half(\Phi^+ + \Phi^-)$ and $\Phi^1= \Phi^{-1}:= \half(\Phi^+ - \Phi^-)$.
The scattering amplitude between particles of types $a$ and $b$ is then
\begin{equation}
  S^{a,b}(\lambda,\lambda') = \exp \left[
    i \int_0^{\lambda-\lambda'} \rmd\nu \ \Phi^{a-b}(\nu)
  \right] \,.
\end{equation}
We give the kernels $\Phi^a$ in Fourier space:
\begin{equation}
  \wh{\Phi}^0(\omega) = \frac{-2\pi \cosh(\pi-3\gamma)\omega}
  {2\sinh \gamma \omega \ \sinh(\pi-2\gamma)\omega} \,,
  \qquad
  \wh{\Phi}^{\pm 1}(\omega) = \frac{2\pi \cosh \gamma \omega}
  {2\sinh \gamma \omega \ \sinh(\pi-2\gamma)\omega} \,.
\end{equation}

\subsection{CFT spectrum}

The CFT describing the
staggered vertex model is a combination of two bosons $\phi_1$ and $\phi_2$, with
action (see \eqref{eq:action-CG})
\begin{equation} \label{eq:action2}
  {\cal A} = \frac{1}{4\pi} \int \rmd^2 x \ \left[
    g_1 (\nabla \phi_1)^2 + g_2 (\nabla \phi_2)^2
  \right] \,,
  \qquad
  \phi_\alpha \equiv \phi_\alpha+2\pi \,,
\end{equation}
where
\begin{equation}
  g_1 = \frac{\gamma}{\pi} \,,
  \qquad
  g_2 \mathop{\longrightarrow}_{N \to \infty} 0 \,.
\end{equation}

\begin{figure}[th]
  \centerline{\psfig{file=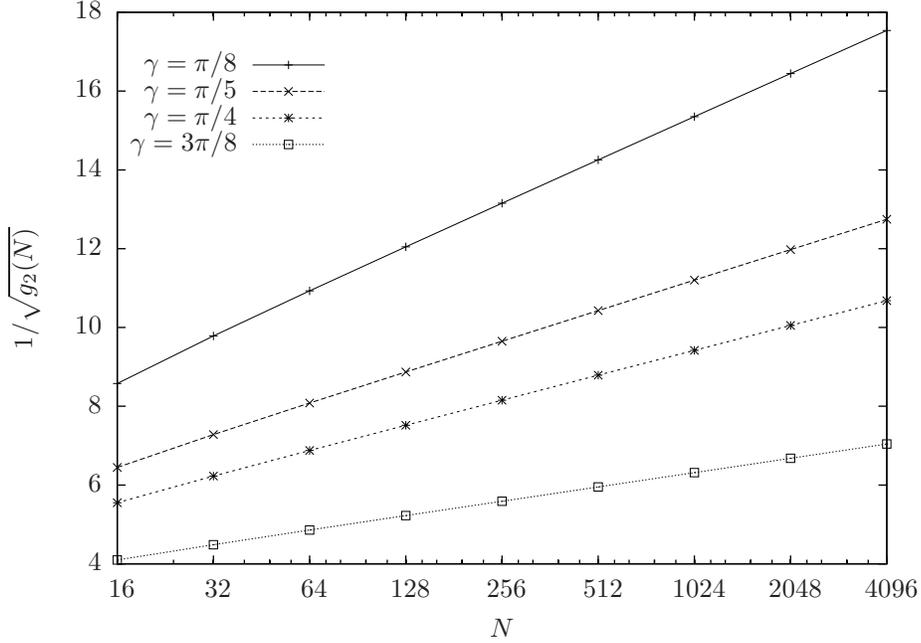, scale=1}}
  \vspace*{8pt}
  \caption{Effective coupling constant of boson $\phi_2$ for the AF critical line, for various
    values of $\gamma$. The data indicate that $g_2(N) \propto \left(\log \frac{N}{N_0} \right)^{-2}$.
  }
  \label{fig:g2}
\end{figure}

As it is done for the XXZ model in Section~\ref{sec:CFT-XXZ}, the above action
is inferred from the central charge and operator content of the BAE.
Since the ground state is of the XXZ type and the Fermi velocity has the XXZ value,
we easily find the central charge of the staggered vertex model:
\begin{equation}
  c = 2 \,.
\end{equation}
Similarly to the XXZ model, the primary operators correspond to higher magnetization
sectors (magnetic operators) and global shifts of the Bethe integers (electric operators).
In the present situation, each set of roots $\{\lambda_j^0\}, \{\lambda_j^1\}$ can be excited
independently, and we denote $m^0,m^1$ ({\resp} $e^0,e^1$) the magnetic ({\resp} electric)
charges. The Wiener-Hopf technique\cite{YY} can be adapted\cite{IJS-PAF-or} to the $\Zbb_2$ BAE,
and we obtain the conformal dimensions
\begin{eqnarray}
  h_{\ul{e},\ul{m}} &=& \frac{1}{8} \left( \frac{e^+}{\sqrt{2g_1}} + m^+ \sqrt{2g_1} \right)^2
  + \frac{1}{8} \left( \frac{e^-}{\sqrt{2g_2}} + m^- \sqrt{2g_2} \right)^2 \,,
  \label{eq:h-PAF} \\
  \hb_{\ul{e},\ul{m}} &=& \frac{1}{8} \left( \frac{e^+}{\sqrt{2g_1}} - m^+ \sqrt{2g_1} \right)^2
  + \frac{1}{8} \left( \frac{e^-}{\sqrt{2g_2}} - m^- \sqrt{2g_2} \right)^2 \,,
  \label{eq:hb-PAF}
\end{eqnarray}
where $e^\pm:= e^0 \pm e^1, m^\pm:= m^0 \pm m^1$, and $g_1,g_2$ are encoded in the scattering kernels:
\begin{equation}
  g_1 = \frac{1}{4} \left[ 1-\frac{\wh{K}^+(0)}{2\pi} \right] \,,
  \qquad
  g_2 = \frac{1}{4} \left[ 1-\frac{\wh{K}^-(0)}{2\pi} \right] \,.
\end{equation}
The above spectrum corresponds exactly to two decoupled bosons
as in~\eqref{eq:action2}. However, the electromagnetic charges
obey the parity conditions
\begin{equation}
  e^+ + e^- \equiv 0 \ {[2]} \,,
  \qquad
  m^+ + m^- \equiv 0 \ {[2]} \,,
\end{equation}
which introduce a coupling of the excitation sectors (or boundary conditions)
for the two bosons. 
The vanishing of $g_2$ makes $\phi_2$ a {\it non-compact}
boson in the continuum limit, and each conformal exponent has a continuum of $m^-$ magnetic
states above it. The level density of this continuum is given by the finite-size scaling of $g_2$.
The latter can be calculated numerically for large but finite $N$, 
by solving the BAE in sectors with $m^- \neq 0$. One gets the following results
\begin{equation}
  g_2(N) \propto \begin{cases}
    \left( \log \frac{N}{N_0} \right)^{-2} & \text{for $0 \leq \gamma < \frac{\pi}{2}$,} \\
    \left( \log \frac{N}{N_0} \right)^{-1} & \text{for $\gamma \to \frac{\pi}{2}$,}
  \end{cases}
\end{equation}
where $N_0$ is a $\gamma$-dependent cut-off value. The numerical results for $g_2$ are shown
in Fig.~\ref{fig:g2}.

\subsection{Effective central charge and critical exponents}

In the presence of a twist $\varphi$, the conformal dimensions~\eqref{eq:h-PAF}--\eqref{eq:h-PAF}
are modified by the change $e^+ \to e^+ + \frac{2\varphi}{\pi}$. The ground state of the critical AF Potts
model has a twist $\varphi=\gamma$, and we denote $e_0:=\frac{\gamma}{\pi}$. The effective central charge
is thus
\begin{equation}
  c_{\rm eff} = 2 - \frac{6 e_0^2}{g_1} = 2 - \frac{6}{t} \,.
\end{equation}

Similarly to Section~\ref{sec:exponents-BK}, some critical exponents of the Potts model
at the AF transition are obtained from various sectors of the $\Zbb_2$ vertex model.
\begin{itemize}

\item The $\ell$-leg watermelon exponent has magnetic charge $m^+=\frac{\ell}{2}$. The
antisymmetric magnetic charge does not contribute to the conformal dimension, since $g_2=0$.
Hence:
\begin{equation}
  X_\ell = \frac{g_1 (\ell/2)^2}{2} - \frac{e_0^2}{2g_1}
  = \frac{\ell^2-4}{8t} \,.
\end{equation}

\item The `thermal sector' with $m^+=0$ contains a state with twist $\varphi = \pi(e_0-1)$,
  and exponent:
  \begin{equation}
    X'_T = \frac{(e_0-1)^2 - e_0^2}{2g_1} = \frac{t-2}{2} \,.
  \end{equation}
  However, the first excited state in the Potts model has an exponent
  \begin{equation}
    X_T = \frac{2(t-3)}{t-2} \,,
  \end{equation}
  which is not part of the electromagnetic spectrum \eqref{eq:h-PAF}--\eqref{eq:h-PAF}.
  This exponent may be interpreted from parafermionic CFTs for integer $t$
  (see Section~\ref{sec:conclusion}), and is valid for generic $t$.

\item In the case of the magnetic exponent $X_H$, the expected value when setting the twist to
  $\varphi=\halfpi$ would be $X_H = \frac{(1/2)^2}{2g_1} - \frac{e_0^2}{2g_1} = \frac{t}{8} - \frac{1}{2t}$, but transfer-matrix
  calculations do not match this value. Rather, the magnetic exponent is
  \begin{equation}
    X_H = \frac{1}{4} - \frac{1}{2t} \,.
  \end{equation}
  Again, this value is consistent with parafermionic CFT for integer $t$.

\end{itemize}

\subsection{The limit $Q \to 0$}
\label{sec:Q-zero}

\begin{figure}[th]
  \centerline{\psfig{file=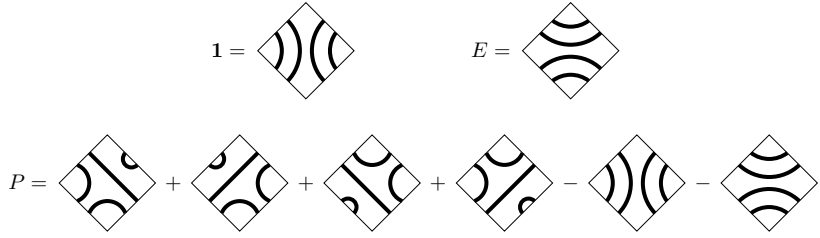, scale=0.8}}
  \vspace*{8pt}
  \caption{Generators of the Brauer model, built from the $\TL_{2N}(n=0)$ algebra.}
  \label{fig:brauer}
\end{figure}

We consider the following limit:
\begin{equation}
  \gamma = \frac{\pi}{2} + h \,,
  \qquad u = h w \,,
  \qquad \text{$w$ fixed} \,, h \to 0 \,.
\end{equation}
In this limit, the double-edge $\Rc$-matrix for the loop model has the
form:
\begin{equation}
  -\frac{\check{\cal R}}{h^2}
  \ \mathop{\longrightarrow}_{h \to 0}
  \ (1-w) \ \id + w \ E + w(1-w) \ P
  \ := \ \check{\cal R}_B(w) \,,
\end{equation}
where $\id$, $E$ and $P$ are defined by the diagrams in Fig.~\ref{fig:brauer}.
The operators $E_j$ and $P_j$ generate a Brauer algebra with loop weight $0$,
defined by the relations
\begin{equation}
  \left\{ \begin{array}{l}
      P_j P_{j \pm 1} P_j = P_{j \pm 1} P_j P_{j \pm 1} \\
      P_j^2 = \id \\
      P_j P_k = P_k P_j \quad \text{for $|j-k|>1$}
    \end{array}\right.
  \qquad
  \left\{ \begin{array}{l}
      E_j E_{j \pm 1} E_j = E_j \\
      E_j^2 = 0 \\
      E_j E_k = E_k E_j \quad \text{for $|j-k|>1$}
    \end{array}\right.
\end{equation}
\begin{equation}
  \left\{ \begin{array}{l}
      P_j E_j = E_j P_j = E_j \\
      E_j P_{j \pm 1} P_j = E_j E_{j \pm 1} \\
      P_{j \pm 1} P_j E_{j \pm 1} = E_j E_{j \pm 1}
    \end{array}\right.
\end{equation}

Moreover, in the above limit, the vertex model is equivalent to the supersymmetric
${\rm OSP}(2|2)$ representation of the Brauer algebra. 
As a check, we can derive the $Q \to 0$ limit of the BAE, where $\lambda_j^a \to h \eta_j^a$:
\begin{equation}
  \left(
    \frac{i+\eta_j^a}{i-\eta_j^a}
  \right)^N
  = (-1)^{r^a-1} e^{2i\varphi} \prod_{\ell=1}^{r^{1-a}}
  \frac{2i+\eta_j^a-\eta_\ell^{1-a}}
  {2i-\eta_j^a+\eta_\ell^{1-a}} \,,
\end{equation}
\begin{equation}
  E = \sum_{a=0,1} \sum_{j=1}^{r^a} \frac{-4}{1+(\eta_j^a)^2} \,,
\end{equation}
which are indeed the ${\rm OSP}(2|2)$ BAE.

Loop models based on the Brauer algebra are integrable, but do not admit a
Coulomb-gas description because loop intersections are present. 
In fact, it was shown\cite{JRS} that they correspond to a universality class distinct from
the $\On$ model. For integer loop weight, they admit a Bethe-Ansatz solution related to the
${\rm OSP}(n|2m)$ superalgebra\cite{OSP-chain}.

\section{The anisotropic critical regime}
\label{sec:PAF-or}

In this Section, we describe the critical regime of the Potts model based on
the $\Zbb_2$ staggered 6V model (like the AF critical line),  in the range $0<u<\gamma$
where the coupling constants $x_1$ and $x_2$ have opposite signs\cite{IJS-PAF-or}
(see Table~\ref{table:regimes}).
We refer to this regime as the anisotropic critical regime. Since it is based on the same
vertex model as the AF critical line, the Bethe-Ansatz analysis is very similar, and so
we will simply emphasize the differences with the previous Section. The resulting CFT
is also a combination of two decoupled bosons $\phi_1$ and $\phi_2$, but in the present case,
both bosons have
a finite compactification radius. It is possible to write the toroidal partition
sum as a generating series of conformal dimensions, and interpret $\phi_2$ as a combination
of two pairs of Majorana fermions.

\subsection{Bethe Ansatz Equations}

The Bethe roots are of two kinds:
\begin{equation}
  \alpha_j^0= \lambda_j^0 \,,
  \qquad
  \alpha_j^1= \lambda_j^1 + i\pi \,,
\end{equation}
and the BAE and energies read
\begin{equation}
  \exp \left[i N k(\lambda_j^a) \right] = (-1)^{r^a-1} e^{2i\varphi}
  \prod_{b=0,1} \prod_{\ell=1}^{r^b}
  \exp \left[-i \Theta^{a-b}(\lambda_j^a-\lambda_\ell^b) \right] \,,
\end{equation}
\begin{equation}
  E = \sum_{a=0,1} \sum_{j=1}^{r^a} \epsilon(\lambda_j^a)
  = \sum_{a=0,1} \sum_{j=1}^{r^a} 
  \frac{-2\sin^2 2\gamma}{\cosh 2\lambda_j^a - \cos 2\gamma}
  \,,
\end{equation}
where:
\begin{equation}
  k(\lambda) := -i \log \frac{\sinh(i\gamma-\lambda)}
    {\sinh(i\gamma+\lambda)} \,,
\end{equation}
\begin{equation}
  \Theta^0(\lambda) := -i \log \frac{\sinh (i\gamma+\frac{\lambda}{2})}
  {\sinh (i\gamma-\frac{\lambda}{2})} \,,
  \qquad
  \Theta^{\pm 1}(\lambda) := -i \log \frac{\cosh (i\gamma+\frac{\lambda}{2})}
  {\cosh (i\gamma-\frac{\lambda}{2})} \,.
\end{equation}

States where $\{ \lambda_j^0 \} = \{ \lambda_j^1 \}$ are solutions of an XXZ model with $\mu=2\gamma$.
The ground state is a double Fermi sea with central charge $c=2$.
Quasi-particles are of two types $0,1$, with dispersion relation~\eqref{eq:dispersion},
and scattering kernels
\begin{equation}
  \wh{\Phi}^0(\omega) = \frac{-2\pi \sinh(\pi-3\gamma)\omega}
  {2 \cosh \gamma \omega \ \sinh(\pi-2\gamma)\omega} \,,
  \qquad
  \wh{\Phi}^{\pm 1}(\omega) = \frac{2\pi \sinh \gamma \omega}
  {2 \cosh \gamma \omega \ \sinh (\pi-2\gamma)\omega} \,.
\end{equation}

\subsection{CFT spectrum and continuum partition function}

The conformal spectrum has the same form~\eqref{eq:h-PAF}--\eqref{eq:hb-PAF} as
for the AF critical point:
\begin{eqnarray}
  h_{\ul{e},\ul{m}} &=& \frac{1}{8} \left( \frac{e^+}{\sqrt{2g_1}} + m^+ \sqrt{2g_1} \right)^2
  + \frac{1}{8} \left( \frac{e^-}{\sqrt{2g_2}} + m^- \sqrt{2g_2} \right)^2 \,,
  \label{eq:h-PAF-or} \\
  \hb_{\ul{e},\ul{m}} &=& \frac{1}{8} \left( \frac{e^+}{\sqrt{2g_1}} - m^+ \sqrt{2g_1} \right)^2
  + \frac{1}{8} \left( \frac{e^-}{\sqrt{2g_2}} - m^- \sqrt{2g_2} \right)^2 \,,
  \label{eq:hb-PAF-or}
\end{eqnarray}
now with finite coupling constants
\begin{equation}
  g_1 = \frac{\pi - 2\gamma}{2\pi} \,, \qquad g_2 = \half \,.
\end{equation}
The parity conditions also hold:
\begin{equation} \label{eq:parity-em}
  e^+ + e^- \equiv 0 \ {[2]} \,,
  \qquad
  m^+ + m^- \equiv 0 \ {[2]} \,.
\end{equation}
In this context, the toroidal partition function can be written as a generating function
for the conformal weights, and it reflects the coupling of the BC:
\begin{equation}
  Z_{\rm stag. \ 6V} = \frac{1}{|\eta(q)|^4}
  \sum_{\myscript{e^+ + e^- \equiv 0 \ [2]}{m^+ + m^- \equiv 0 \ [2]}}
  q^{h_{\ul{e},\ul{m}}} \ \qb^{\hb_{\ul{e},\ul{m}}} \,.
\end{equation}
The factors associated to charges $e^-,m^-$ are bosonic partition sums with coupling
constant $g_2=\half$, which can be written as products of {\it two} Ising partition sums
with various BC. We introduce the Ising partition sum ${\cal Z}_{r,r'}$, with BC on the Ising
spins $\sigma \to (-1)^r \sigma, \sigma \to (-1)^{r'} \sigma$, and the bosonic partition sum $Z_{m,m'}(g)$, with
BC $\phi \to \phi + 2\pi m, \phi \to \phi + 2\pi m'$. The partition function reads\cite{IJS-PAF-or}
\begin{equation}
  Z_{\rm stag. \ 6V} =
  \half \sum_{\myscript{m \equiv r_1+r_2 \ {[2]}}{m' \equiv r'_1+r'_2 \ {[2]}}}
  (-1)^{r_1 r'_2 + r'_1 r_2} \ {\cal Z}_{r_1,r'_1} \ {\cal Z}_{r_2,r'_2}
  \ Z_{m,m'}(g_1) \,.
\end{equation}
In this form, the $\Zbb_2$ model appears as two Ising models and one compact boson,
decoupled in the bulk and coupled through their BC.

\subsection{Effective central charge and critical exponents}

In the presence of a twist $\varphi$, the conformal dimensions~\eqref{eq:h-PAF-or}--\eqref{eq:h-PAF-or}
are modified by the change $e^+ \to e^+ + \frac{2\varphi}{\pi}$. Like for the AF line,
we denote $e_0:=\frac{\gamma}{\pi}$, and the ground state has a twist $\varphi=\pi e_0$.
The effective central charge is thus
\begin{equation}
  c_{\rm eff} = 2 - \frac{6 e_0^2}{g_1} = 2 - \frac{12}{t(t-2)} \,.
\end{equation}

\begin{itemize}

\item The $\ell$-leg watermelon exponent has magnetic charge $m^+=\frac{\ell}{2}$. Because of the
parity condition~\eqref{eq:parity-em}, the lowest possible value for the antisymmetric magnetic
charge $m^-$ is $0$ for even $m^+$ and $1$ for odd $m^+$. As a consequence, the watermelon
exponents are
\begin{equation}
  X_\ell = \begin{cases}
    \frac{g_1 (\ell/2)^2}{2} - \frac{e_0^2}{2g_1} & \text{if $\ell \equiv 0 \ {[4]}$} \\
    \frac{g_1 (\ell/2)^2}{2} + \frac{1}{4} - \frac{e_0^2}{2g_1} & \text{if $\ell \equiv 2 \ {[4]}$} \,.
  \end{cases}
\end{equation}

\item In the thermal sector with twist $\varphi=e_0$, the energy exponent corresponds to
  charges $e^+ = 2e_0-2, e^-=0$, which gives the constant value
  \begin{equation}
    X_T = \frac{(e_0-1)^2 - e_0^2}{2g_1} = 1 \,.
  \end{equation}
  This is equivalent to the energy excitation in one of the Ising models.

\item The magnetic exponent $X_H$ is given by the twist $\varphi=\halfpi$. It turns out that the lowest
  exponent in this twist sector has electric charges $e^+= \frac{2\varphi}{\pi}-1, e^-= 1$. The exponent is thus
  \begin{equation}
    X_H = \frac{1}{4} - \frac{e_0^2}{2g_1} \,.
  \end{equation}

\end{itemize}

\section{Conclusion}
\label{sec:conclusion}

The $Q$-state Potts model is a simple lattice model, in which the ferromagnetic
critical point provides a basic realization of minimal CFTs.
The {\af} part of its phase diagram has been less studied than the
ferromagnetic one, but it contains three critical regimes of physical interest for
Statistical Mechanics and Condensed Matter Theory.
In this review, we have shown that, through an equivalence to loop and vertex models,
these critical points can be solved by Bethe Ansatz. This approach is powerful enough
to discover the CFT spectrum, the $S$-matrix description of quasi-excitations, and
also the critical exponents in the original statistical model.

Most of the material exposed in this review is valid for generic values of
$Q$. When $Q$ is a Beraha number ($Q=4 \cos^2 \frac{\pi}{t}$, with $t$ integer),
the Potts model can be mapped to an RSOS height model
(see Section~\ref{sec:related}). Although the partition functions of the two models
are identical (even at the lattice level), the excitation spectra can be very different.
The essential mechanism at play is that some excited states of the vertex model have
a vanishing contribution to the RSOS partition function. In the CFT language, the elimination
of these states amounts to {\it null-state equations}, which are the basis for the determination
of minimal series of CFTs. RSOS models with homogeneous Boltzmann weights (corresponding to
the self-dual Potts model) were shown to have four different physical phases I,II,III,IV, with
critical transitions separating phases III-IV and I-II. These two transitions are lattice
realizations of the Minimal and $\Zbb_k$-parafermionic (for $k=t-2$) CFTs, respectively\cite{Huse}.

Back to the Potts model with Beraha values, this tells us that the RSOS version of
the ferromagnetic critical
point relates to a Minimal CFT, whereas that of the BK phase relates to the 
$\Zbb_k$-parafermionic CFT. The question of the RSOS version for the AF critical line is
more difficult, but it has been found recently\cite{JS-PAF} that this is also described
by the $\Zbb_k$-parafermionic CFT. This relation is supported by the study of
critical exponents and of the algebra of chiral currents.

Although the critical phases of the {\af} Potts model are now identified and well studied,
there remain interesting open questions to the mathematical physicist.

From the point of view of integrable models, 
the understanding of the $\Zbb_2$ model would be more complete if one could
extend the relation to braiding algebras (see Section~\ref{sec:Q-zero}).
Also, an analytic derivation of the effective coupling constant for the non-compact boson
should be possible. Moreover, one can generalize the $\Zbb_2$ construction to a $\Zbb_p$ six-vertex
model, where the $\Zbb_p$ symmetry will be reflected in the whole Bethe-Ansatz structure.
This opens a new series of models, similarly to what appears in the fusion procedure.

Finally, the Potts model is a good framework to study the physics of cluster or
spin interfaces in lattice models. Current studies focus on the links between boundary CFT,
Schramm-L\"owner Evolution (SLE) and free massless field theories.
In the case of the {\af} Potts model, the Fortuin-Kasteleyn cluster
boundaries are well described by the $\ell$-leg watermelon exponents, all related to the
first boson $\phi_1$, but the geometric expression of the second boson $\phi_2$ is lacking.
This additional degree of freedom may lead to an extended version of the SLE process,
as was introduced recently\cite{Bettelheim,Raoul}.

\section*{Acknowledgements}

The author thanks J. Jacobsen and H. Saleur for their help with the preparation
of this paper.

\end{document}